\documentclass[a4paper]{article}

\usepackage{INTERSPEECH2020}

\usepackage{hyperref}
\usepackage{url}
\usepackage{bm}
\usepackage{multirow}
\usepackage{tabularx}
\newcolumntype{C}{>{\centering\arraybackslash}X}
\makeatletter
\newcommand\footnoteref[1]{\protected@xdef\@thefnmark{\ref{#1}}\@footnotemark}
\makeatother

\title{CycleGAN-VC3:\\
  Examining and Improving CycleGAN-VCs for Mel-spectrogram Conversion}
\name{Takuhiro Kaneko, Hirokazu Kameoka, Kou Tanaka, Nobukatsu Hojo}
\address{
  NTT Communication Science Laboratories, NTT Corporation, Japan}
\email{takuhiro.kaneko.tb@hco.ntt.co.jp}

\begin{document}

\maketitle
\begin{abstract}
  Non-parallel voice conversion (VC) is a technique for learning mappings between source and target speeches without using a parallel corpus. Recently, cycle-consistent adversarial network (CycleGAN)-VC and CycleGAN-VC2 have shown promising results regarding this problem and have been widely used as benchmark methods. However, owing to the ambiguity of the effectiveness of CycleGAN-VC/VC2 for mel-spectrogram conversion, they are typically used for mel-cepstrum conversion even when comparative methods employ mel-spectrogram as a conversion target. To address this, we examined the applicability of CycleGAN-VC/VC2 to mel-spectrogram conversion. Through initial experiments, we discovered that their direct applications compromised the time-frequency structure that should be preserved during conversion. To remedy this, we propose CycleGAN-VC3, an improvement of CycleGAN-VC2 that incorporates time-frequency adaptive normalization (TFAN). Using TFAN, we can adjust the scale and bias of the converted features while reflecting the time-frequency structure of the source mel-spectrogram. We evaluated CycleGAN-VC3 on inter-gender and intra-gender non-parallel VC. A subjective evaluation of naturalness and similarity showed that for every VC pair, CycleGAN-VC3 outperforms or is competitive with the two types of CycleGAN-VC2, one of which was applied to mel-cepstrum and the other to mel-spectrogram.\footnote{\label{foot:samples}Audio samples are available at \url{http://www.kecl.ntt.co.jp/people/kaneko.takuhiro/projects/cyclegan-vc3/index.html}.}
\end{abstract}

\section{Introduction}
\label{sec:introduction}

Voice conversion (VC) is a technique for converting non/para-linguistic information in speech while retaining the linguistic information.
VC has been actively studied owing to its potential diverse applications such as in speaking aids~\cite{AKainSC2007,KNakamuraSC2012}, speech enhancement~\cite{ZInanogluSC2009,TTodaTASLP2012}, and accent conversion~\cite{TKanekoIS2017b}.
Machine-learning approaches are widely used and include statistical methods based on Gaussian mixture models~\cite{YStylianouTASP1998,TTodaTASLP2007} and neural networks (NNs), including feedforward NNs~\cite{SDesaiTASLP2010}, recurrent NNs~\cite{LSunICASSP2015}, convolutional NNs (CNNs)~\cite{TKanekoIS2017b}, attention networks~\cite{KTanakaICASSP2019,HKameokaTASLP2020}, and generative adversarial networks (GANs)~\cite{TKanekoIS2017b}.

For ease of learning, many VC methods (including the above-mentioned methods) learn mappings from source to target speeches using a parallel corpus.
However, collection of such data is not necessarily easy or practical.
Additionally, even if such data are collected, the time alignment procedure required in most VC methods remains a challenge.

\begin{figure}[t]
  \centering
  \includegraphics[width=0.99\columnwidth]{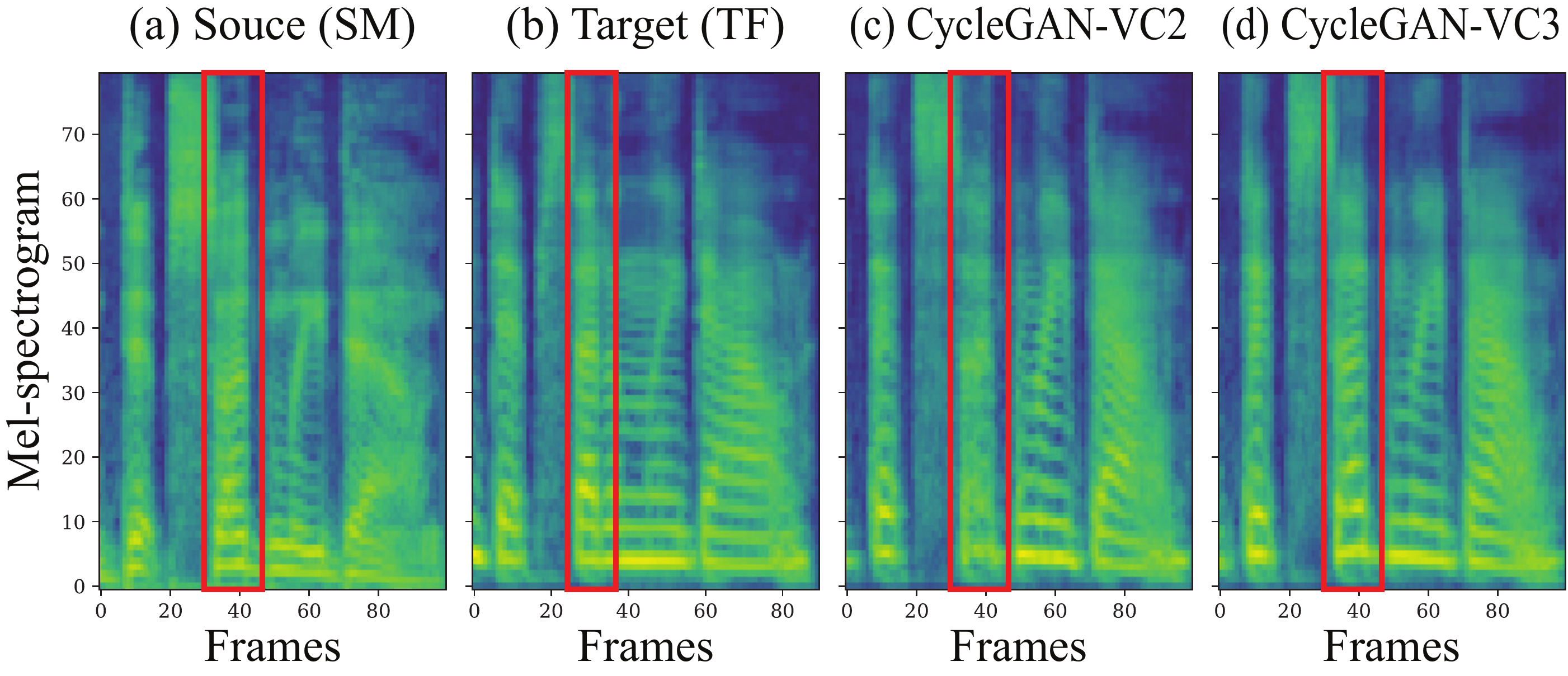}
  \vspace{-2mm}
  \caption{
    Comparison of source, target, and converted mel-spectrograms.
    In the red box, CycleGAN-VC2 (c) compromises the harmonic structure that appears in the source (a) and target (b), whereas CycleGAN-VC3 (d) succeeds in retaining it.
    Note that achieving this is not trivial because parallel data, such as (a) and (b), were not available for training.
  }
  \vspace{-5mm}
  \label{fig:mel-spectrogram}
\end{figure}

As an alternative, non-parallel VC, which does not require a parallel corpus for training, has gained attention recently.
Non-parallel VC is attractive in terms of data collection cost; however, its learning is challenging owing to the absence of explicit supervision.
To address this, several studies have utilized linguistic information~\cite{LSunICME2016,FLXieIS2016,YSaitoICASSP2018,JZhangTASLP2020}.
Although this extra supervision improves the performance, auxiliary data or modules are required to extract linguistic information.

To avoid such a requirement and achieve non-parallel VC using only acoustic data, variational autoencoder-based methods~\cite{CHsuAPSIPA2016,CHsuIS2017,HKameokaTASLP2019} and GAN-based methods~\cite{CHsuIS2017,TKanekoArXiv2017} have been proposed.
Among them, CycleGAN-VC~\cite{TKanekoArXiv2017,TKanekoEUSIPCO2018} has garnered attention alongside its variants (CycleGAN-VC2~\cite{TKanekoICASSP2019} and StarGAN-VCs~\cite{HKameokaSLT2018,TKanekoIS2019,HKameokaArXiv2020}), and they have been widely used as benchmark methods in several studies (e.g.,~\cite{JZhangTASLP2020,SLeeICASSP2020,KQianICML2019}).
However, owing to the ambiguity of their effectiveness for mel-spectrogram conversion, they are typically used for mel-cepstrum conversion even when comparative methods employ mel-spectrogram as a conversion target (e.g., \cite{JZhangTASLP2020,KQianICML2019}).

These facts motivated us to examine the applicability of CycleGAN-VC~\cite{TKanekoArXiv2017} and CycleGAN-VC2~\cite{TKanekoICASSP2019} to mel-spectrogram conversion.
Through initial experiments, we found that when CycleGAN-VC/VC2 is directly applied to a mel-spectrogram, it compromises the time-frequency structure that should be preserved during conversion, as shown in Figure~\ref{fig:mel-spectrogram}.
To address this, we propose CycleGAN-VC3 that is an improvement of CycleGAN-VC2 that incorporates \textit{time-frequency adaptive normalization (TFAN)}.
TFAN is inspired by spatially adaptive (de)normalization (SPADE)~\cite{TParkCVPR2019} that was originally proposed for semantic image synthesis.
We revise SPADE for application to 1D and 2D time-frequency features.
Using TFAN, we can adjust the scale and bias of the converted features while reflecting the time-frequency structure of the source mel-spectrogram.

We examined the effectiveness of the CycleGAN-VC3 on inter-gender and intra-gender non-parallel VC using the Voice Conversion Challenge 2018 (VCC 2018) dataset~\cite{VCC2018}.
A subjective evaluation of naturalness and similarity showed that for every VC pair, CycleGAN-VC3 achieved a better or competitive performance compared with the two types of CycleGAN-VC2, one of which was applied to mel-cepstrum and the other to mel-spectrogram.

The remainder of this paper is organized as follows.
In Section~\ref{sec:cycleganvc-vc2}, we briefly review conventional CycleGAN-VC/VC2.
In Section~\ref{sec:cycleganvc3}, we introduce the proposed CycleGAN-VC3.
In Section~\ref{sec:experiments}, we describe the experimental results.
Section~\ref{sec:conclusions} presents a concise summary and description of future work.

\section{Conventional CycleGAN-VC/VC2}
\label{sec:cycleganvc-vc2}

\subsection{Training objectives}
\label{subsec:objective}

CycleGAN-VC/VC2 aims at learning a mapping $G_{X \rightarrow Y}$ that converts source acoustic features $\bm{x} \in X$ into target acoustic features $\bm{y} \in Y$ without using a parallel corpus.
Inspired by CycleGAN~\cite{JYZhuICCV2017}, originally proposed for unpaired image-to-image translation, CycleGAN-VC/VC2 learns a mapping using an \textit{adversarial loss}~\cite{IGoodfellowNIPS2014}, \textit{cycle-consistency loss}~\cite{TZhouCVPR2016}, and \textit{identity-mapping loss}~\cite{YTaigmanICLR2017}.
Furthermore, CycleGAN-VC2 uses a \textit{second adversarial loss} to improve the details of the reconstructed features.

\smallskip\noindent\textbf{Adversarial loss.}
To ensure that the converted feature $G_{X \rightarrow Y} (\bm{x})$ is in the target $Y$, an adversarial loss ${\cal L}_{adv}^{X \rightarrow Y}$ is used as follows:
\begin{flalign}
  \label{eqn:adv}
  {\cal L}_{adv}^{X \rightarrow Y}
  = & \: \mathbb{E}_{\bm{y} \sim P_Y} [\log D_Y(\bm{y})]
  \nonumber \\
  + & \: \mathbb{E}_{\bm{x} \sim P_X} [\log (1 - D_Y(G_{X \rightarrow Y}(\bm{x})))],
\end{flalign}
where the discriminator $D_{Y}$ attempts to distinguish synthesized $G_{X \rightarrow Y} (\bm{x})$ from real $\bm{y}$ by maximizing the loss, whereas $G_{X \rightarrow Y}$ attempts to synthesize $G_{X \rightarrow Y} (\bm{x})$ that can deceive $D_{Y}$ by minimizing the loss.
Similarly, the inverse mapping $G_{Y \rightarrow X}$ and the discriminator $D_{X}$ are trained adversarially using ${\cal L}_{adv}^{Y \rightarrow X}$.

\smallskip\noindent\textbf{Cycle-consistency loss.}
To preserve the composition in conversion, a cycle-consistency loss ${\cal L}_{cyc}$ is used as follows:
\begin{flalign}
  \label{eqn:cycle}
  {\cal L}_{cyc}
  = & \: \mathbb{E}_{\bm{x} \sim P_X}[\| G_{Y \rightarrow X}(G_{X \rightarrow Y}(\bm{x})) - \bm{x} \|_1 ]
  \nonumber \\
  + & \: \mathbb{E}_{\bm{y} \sim P_Y}[\| G_{X \rightarrow Y}(G_{Y \rightarrow X}(\bm{y})) - \bm{y} \|_1 ].
\end{flalign}
This loss is used with a hyper-parameter $\lambda_{cyc}$, which controls its relative importance.
The loss aids $G_{X \rightarrow Y}$ and $G_{Y \rightarrow X}$ to identify pseudo pairs within the cycle-consistency constraint.

\smallskip\noindent\textbf{Identity-mapping loss.}
To facilitate input preservation, an identity-mapping loss ${\cal L}_{id}$ is used as follows:
\begin{flalign}
  \label{eqn:identity}
  {\cal L}_{id}
  = & \: \mathbb{E}_{\bm{y} \sim P_Y} [\| G_{X \rightarrow Y}(\bm{y}) - \bm{y} \|_1]
  \nonumber \\
  + & \: \mathbb{E}_{\bm{x} \sim P_X} [\| G_{Y \rightarrow X}(\bm{x}) - \bm{x} \|_1].
\end{flalign}
This loss is used with a hyper-parameter $\lambda_{id}$, which controls its relative importance.

\smallskip\noindent\textbf{Second adversarial loss.}
In CycleGAN-VC2, to mitigate the statistical averaging caused by the L1 loss (Equation~\ref{eqn:cycle}), an additional discriminator $D'_{X}$ is introduced, and a second adversarial loss ${\cal L}_{adv2}^{X \rightarrow Y \rightarrow X}$ is imposed on the circularly converted features as follows:
\begin{flalign}
  \label{eqn:adv2}
  &{\cal L}_{adv2}^{X \rightarrow Y \rightarrow X}
  = \mathbb{E}_{\bm{x} \sim P_X} [\log D'_X(\bm{x})]
  \nonumber \\
  &\:\:\:\:\:\:\:\:\:\:\:\:\:
  + \mathbb{E}_{\bm{x} \sim P_X} [\log (1 - D'_X(G_{Y \rightarrow X}(G_{X \rightarrow Y}(\bm x))))].
\end{flalign}
Similarly, the discriminator $D'_{Y}$ is introduced, and ${\cal L}_{adv2}^{Y \rightarrow X \rightarrow Y}$ is imposed on the inverse-forward mapping.

\subsection{Generator architectures}
\label{subsec:generator}

CycleGAN-VC uses \textit{1D CNN generators}~\cite{TKanekoIS2017b} to capture the overall relationship along with the feature direction while preserving the temporal structure.
In particular, the network is composed of downsampling, residual~\cite{KHeCVPR2016}, and upsampling blocks to capture wide-range temporal relationships effectively, and gated linear units (GLUs)~\cite{YDauphinICML2017} are used as activation to learn sequential and hierarchical structure adaptively.

However, a study on CycleGAN-VC2~\cite{TKanekoICASSP2019} showed that 1D CNNs in downsampling and upsampling blocks affect the structure that should be retained in conversion.
To alleviate this, CycleGAN-VC2 introduces \textit{2-1-2D CNN} that uses 2D CNNs in upsampling and downsampling blocks and uses 1D CNNs in residual blocks.
The former is used for extracting the time-frequency structure while preserving the original structure.
The latter is used for performing dynamic changes.

\subsection{Discriminator architectures}
\label{subsec:discriminator}

CycleGAN-VC uses \textit{2D CNN discriminators}~\cite{TKanekoICASSP2017} to discriminate data based on 2D spectral textures.
In particular, it uses \textit{FullGAN}, which has a fully connected layer as the last layer, to discriminate data based on the overall input structure.
However, in FullGAN, the need to learn many parameters causes learning difficulty. To mitigate this, CycleGAN-VC2 introduces \textit{PatchGAN}~\cite{CLiECCV2016} that uses convolution at the last layer.
This reduces the parameters and stabilizes the GAN training.

\section{CycleGAN-VC3}
\label{sec:cycleganvc3}

\subsection{TFAN: Time-frequency adaptive normalization}
\label{subsec:tfan}

CycleGAN-VC and CycleGAN-VC2 were originally designed for mel-cepstrum conversion, and their effectiveness in mel-spectrogram conversion is not sufficiently examined.
We empirically examined their effectiveness and discovered that they compromise the time-frequency structure that should be preserved in conversion, as shown in Figure~\ref{fig:mel-spectrogram}.

\begin{figure}[t]
  \centering
  \includegraphics[width=0.99\columnwidth]{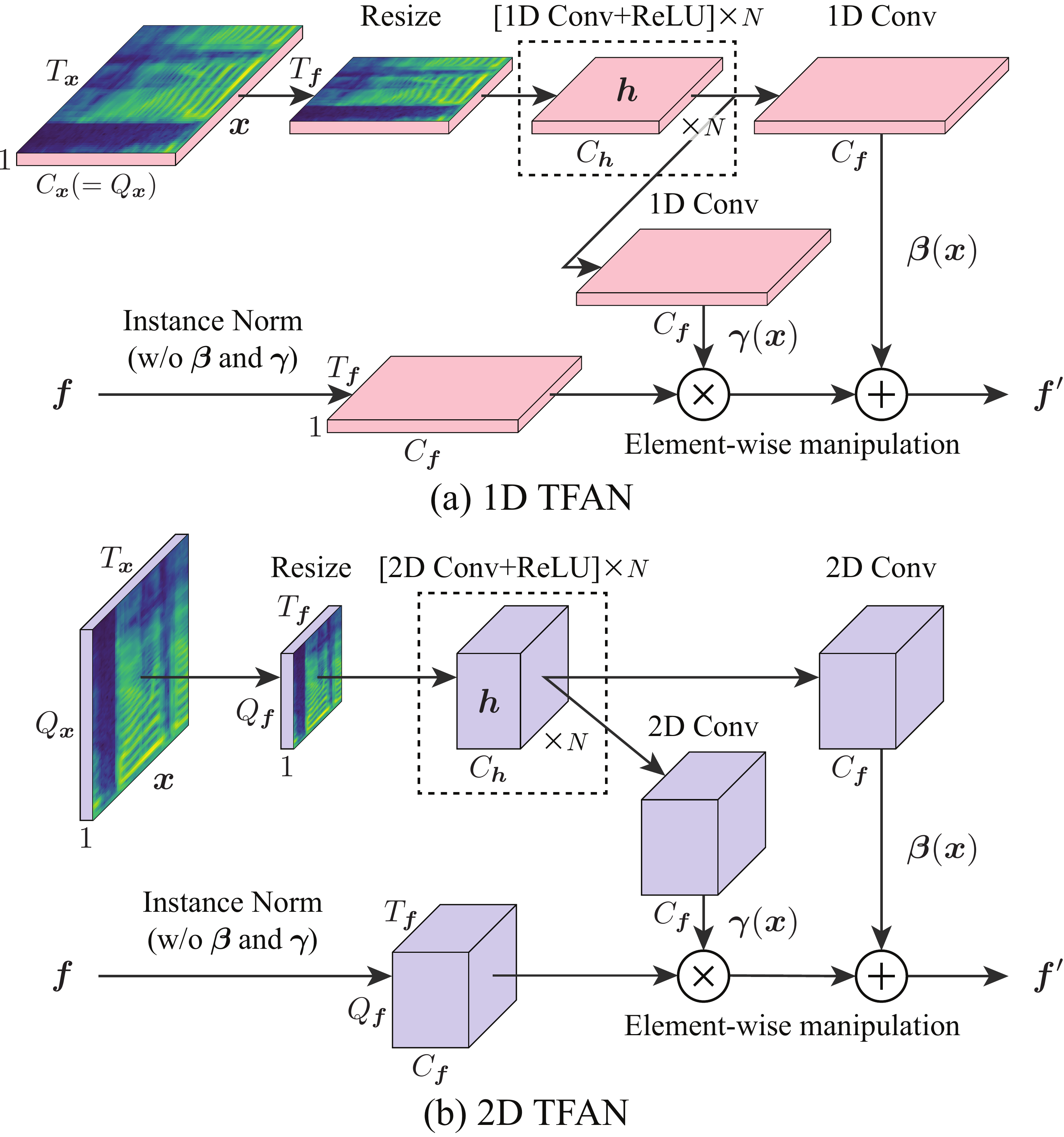}
  \vspace{-2mm}
  \caption{
    Architectures of 1D and 2D TFAN.
    $\bm{x}$, $\bm{f}$, $\bm{f}'$, and $\bm{h}$ indicate the source mel-spectrogram, input feature, output feature, and hidden-layer feature, respectively.
    $T$, $Q$, and $C$ with subscripts denote the time, frequency, and channel dimensions in the corresponding feature, respectively.
    We omit them when their values are the same as those in the previous step.
    $N$ represents the number of layers in the dashed-line box.
    After $\bm{f}$ is normalized in a channel-wise manner, it is modulated in an element-wise manner using scale $\bm{\gamma}({\bm{x}})$ and bias $\bm{\beta}({\bm{x}})$, which are calculated using either (a) 1D CNN or (b) 2D CNN.
  }
  \vspace{-5mm}
  \label{fig:tfan}
\end{figure}

Motivated by this finding, we devised TFAN that extends instance normalization (IN)~\cite{DUlyanovArXiv2016} to adjust the scale and bias of the converted features while reflecting the source information (i.e., $\bm{x}$) in a time- and frequency-wise manner.
In particular, we designed TFAN for 1D and 2D time-frequency features to be used in 2-1-2D CNN (Section~\ref{subsec:generator}).
Figure~\ref{fig:tfan} illustrates the architectures of TFAN.
Given feature $\bm{f}$, TFAN normalizes it in a \textit{channel-wise manner} similar to IN and then modulates the normalized feature in an \textit{element-wise manner} using scale $\bm{\gamma}(\bm{x})$ and bias $\bm{\beta}(\bm{x})$, which are calculated from $\bm{x}$ using CNNs:
\begin{flalign}
  \label{eqn:tfan}
  \bm{f}' = \bm{\gamma}(\bm{x}) \frac{\bm{f} - \bm{\mu}(\bm{f})}{\bm{\sigma}(\bm{f})} + \bm{\beta}(\bm{x}),
\end{flalign}
where $\bm{f}'$ is the output feature, and $\bm{\mu}(\bm{f})$ and $\bm{\sigma}(\bm{f})$ are the channel-wise average and standard deviation of $\bm{f}$, respectively.
In IN, \textit{$\bm{x}$-independent} scale $\bm{\beta}$ and bias $\bm{\gamma}$ are applied in a \textit{channel-wise manner}, whereas in TFAN, those \textit{calculated from $\bm{x}$} (i.e., $\bm{\beta}(\bm{x})$ and $\bm{\gamma}(\bm{x})$) are applied in an \textit{element-wise manner}.
These differences allow TFAN to adjust the scale and bias of $\bm{f}$ while reflecting $\bm{x}$ in a time- and frequency-wise manner.

Note that TFAN is inspired by SPADE~\cite{TParkCVPR2019}, which was originally proposed for semantic image synthesis.
The main differences are that (1) SPADE was devised for 2D image features, whereas TFAN was designed for both 1D and 2D time-frequency features, (2) SPADE uses a one-layer CNN in the component shown in the dashed-line box in Figure~\ref{fig:tfan} because the drastic changes are not required in semantic image synthesis, whereas TFAN uses a multi-layer CNN to ensure the dynamic change, and (3) SPADE is based on batch normalization~\cite{SIoffeICML2015}, whereas TFAN is based on IN.
We examine the effects of (1) and (2) and present our findings in Section~\ref{subsec:objective_evaluation}.

\begin{figure}[b]
  \centering
  \vspace{-5mm}
  \includegraphics[width=0.99\columnwidth]{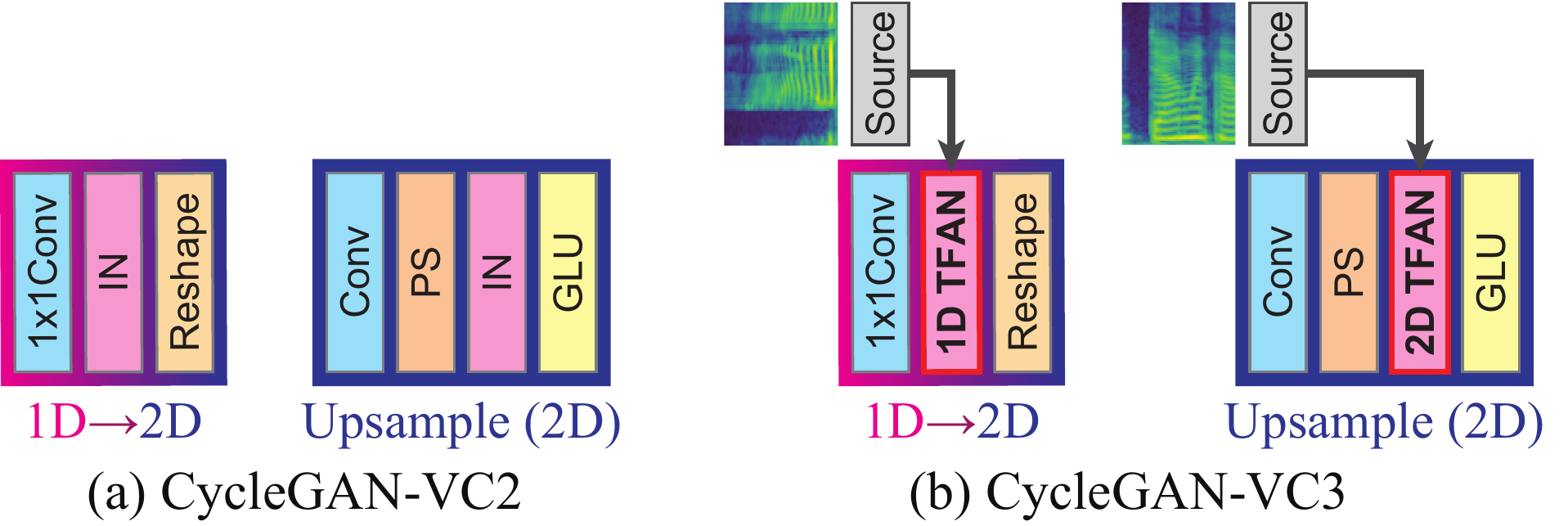}
  \vspace{-2mm}
  \caption{
    Comparison of 1D$\rightarrow$2D and upsampling blocks between (a) CycleGAN-VC2 and (b) CycleGAN-VC3.
    See Figure 4 in \cite{TKanekoICASSP2019} for details of the overall network architectures.
  }
  \label{fig:implementation}
\end{figure}

\subsection{Implementation}
\label{subsec:implementation}

In CycleGAN-VC3, we incorporated TFAN into the CycleGAN-VC2 generator (i.e., 2-1-2D CNN (Section~\ref{subsec:generator})).
Particularly, IN in the 1D$\rightarrow$2D block and that in the upsampling block were replaced with 1D TFAN and 2D TFAN, respectively, as shown in Figure~\ref{fig:implementation}.
In TFAN, we set the number of channels (i.e., $C_{\bm{h}}$) and kernel size in $\bm{h}$ (Figure~\ref{fig:tfan}) to 128 and 5, respectively.
We examine the performance with varying $N$ and adjusting the position where the TFAN is inserted and present our findings in Section~\ref{subsec:objective_evaluation}.
The discriminator was the same as that used in CycleGAN-VC2 (i.e., PatchGAN (Section~\ref{subsec:discriminator})).

\section{Experiments}
\label{sec:experiments}

\subsection{Experimental conditions}
\label{subsec:experimental_conditions}

\textbf{Dataset.}
We evaluated CycleGAN-VCs on the Spoke (i.e., non-parallel VC) task of VCC 2018~\cite{VCC2018}, which contains recordings of professional US English speakers.
We selected a subset of speakers that considered all inter-gender and intra-gender VC: VCC2SF3 (\textit{SF}), VCC2SM3 (\textit{SM}), VCC2TF1 (\textit{TF}), and VCC2TM1 (\textit{TM}), where \textit{S}, \textit{T}, \textit{F}, and \textit{M} represent source, target, female, and male, respectively.
Combinations of 2 sources $\times$ 2 targets were used for evaluation.
For each speaker, 81 utterances (approximately 5 min, which is relatively low for VC) and 35 utterances were used for training and evaluation, respectively.
In the training set, there were no overlaps between the source and target utterances; therefore, this problem must be solved in a fully non-parallel setting.
The recordings were downsampled to 22.05 kHz.
Following the study of MelGAN~\cite{KKumarNeurIPS2019}, which we used as a vocoder in our experiments, we extracted an 80-dimensional log mel-spectrogram with a window length of 1024 and hop length of 256 samples.

\smallskip\noindent\textbf{Conversion process.}
One aim in this study was to examine the feasibility of using CycleGAN-VCs for mel-spectrogram conversion.
Hence, we used CycleGAN-VCs for mel-spectrogram conversion and synthesized waveforms using the pretrained MelGAN vocoder~\cite{KKumarNeurIPS2019}.\footnote{\label{foot:melgan}\url{https://github.com/descriptinc/melgan-neurips}}
We did not alter the parameters of the vocoder such that we could focus on the evaluation of mel-spectrogram conversion; however, fine-tuning them for each speaker is a possible means for improvement.

\smallskip\noindent\textbf{Network architectures.}
As the acoustic feature is changed from mel-cepstrum to mel-spectrogram, the feature dimension increased from 35 to 80.
However, the generators of CycleGAN-VCs are fully convolutional; therefore, they can be used without modifying the network architecture.
Regarding the discriminators, we used the same network architecture as those for mel-cepstrum conversion, except that in CycleGAN-VC2/VC3, the kernel size in the second-last convolutional layer was doubled in the frequency direction (see Figure 4 in \cite{TKanekoICASSP2019} for details of the original network architectures).

\smallskip\noindent\textbf{Training settings.}
The training settings were similar to those used in CycleGAN-VC/VC2 for mel-cepstrum conversion~\cite{TKanekoArXiv2017,TKanekoICASSP2019}.
For preprocessing, we normalized the mel-spectrograms using the mean and variance of the training data.
We used the least square GAN~\cite{XMaoICCV2017} as the GAN objective.
We trained the networks for $500k$ iterations using the Adam optimizer~\cite{DPKingmaICLR2015} with a batch size of 1.
A training sample consisted of randomly cropped 64 frames (approximately 0.75 s).
The learning rates were set to 0.0002 for the generators and 0.0001 for the discriminators with momentum terms $\beta_1$ and $\beta_2$ of 0.5 and 0.999, respectively.
$\lambda_{cyc}$ and $\lambda_{id}$ were set to 10 and 5, respectively, and ${\cal L}_{id}$ was used only for the first $10k$ iterations.
Note that similar to the original CycleGAN-VC/VC2, \textit{we did not use extra data, modules, or time alignment procedures for training}.

\vspace{-1.3mm}
\subsection{Objective evaluation}
\label{subsec:objective_evaluation}

We conducted an objective evaluation to investigate the effect of TFAN parameter selection and the performance difference among CycleGAN-VCs.
Direct measurement of the difference between the target and converted mel-spectrograms is difficult because their alignment is not trivial.
As an alternative, we used two evaluation metrics that are commonly used in previous VC~\cite{TKanekoICASSP2019,TKanekoIS2019,SLeeICASSP2020}: the \textit{mel-cepstral distortion (MCD)}, which measures the global structural difference based on the target and converted mel-cepstra, and \textit{modulation-spectra distance (MSD)}, which assesses the local structural difference based on the target and converted modulation spectra of mel-cepstra.
For both metrics, the smaller the value, the better the performance.
35-dimensional mel-cepstrum parameters were extracted from the target or converted waveform using WORLD~\cite{MMoriseIEICE2016}.

\smallskip\noindent\textbf{Effect of TFAN parameter selection.}
We initially examined the effect of the TFAN parameter selection.
In particular, we investigated the performance with varying the depth of TFAN ($N$ in Figure~\ref{fig:tfan}) and the position where the TFAN is inserted (1D$\rightarrow$2D and/or upsampling blocks (see Figure~\ref{fig:implementation})).
Table~\ref{tab:objective_evaluation}(a) and (b) list the respective results.
Our major findings are as follows.
\textit{(1) Comparison of different-depth TFAN (Table~\ref{tab:objective_evaluation}(a)).}
We found that (i) 1-depth TFAN shows the worst scores except for SF-TF, where the performance is comparable with that of the others, and (ii) the scores reach their peak at approximately $N = 3$.
This indicates the importance of performing dynamic changes using a multi-layer CNN.
This differs from semantic image synthesis with SPADE~\cite{TParkCVPR2019} applied.
\textit{(2) Comparison of TFAN positions (Table~\ref{tab:objective_evaluation}(b)).}
We found that the joint usage of 1D TFAN and 2D TFAN in 1D$\rightarrow$2D and upsampling blocks is the most effective.
Therefore, we set $N$ to 3 and used TFAN in both positions in the following experiments.

\begin{table}[t]
  \caption{
    Comparison of MCD and MSD using (a) different-depth TFAN, (b) TFAN in different positions, and (c) different models.
    The results are listed as MCD [dB]/MSD [dB], where bold and italic numbers indicate the best and second-best scores, respectively.
  }
  \vspace{-2mm}
  \label{tab:objective_evaluation}
  \centering
  \scriptsize{
  \begin{tabularx}{\columnwidth}{cCCCC}
    \toprule
    (a) Depth ($N$) & SF-TF & \!\!SM-TM\!\! & SF-TM & SM-TF
    \\ \midrule
    1
    & 7.55/\textbf{1.54}
    & 7.39/1.83
    & 7.61/1.77
    & 9.48/1.81
    \\
    2
    & \textbf{7.53}/\textit{1.55}
    & 7.23/1.79
    & \textbf{6.91}/\textit{1.76}
    & \textbf{7.97}/\textit{1.61}
    \\
    3
    & \textit{7.54}/\textit{1.55}
    & \textbf{7.10}/\textbf{1.78}
    & \textbf{6.91}/\textbf{1.74}
    & \textbf{7.97}/\textit{1.61}
    \\
    4
    & \textit{7.54}/\textit{1.55}
    & \textbf{7.10}/\textit{1.80}
    & 6.92/1.77
    & 8.01/\textbf{1.59}
    \\ \bottomrule
    \toprule
    (b) Position & SF-TF & \!\!SM-TM\!\! & SF-TM & SM-TF
    \\ \midrule
    1D$\rightarrow$2D
    & 7.79/\textit{1.58}
    & \textit{7.12}/1.93
    & 7.05/1.82
    & \textit{8.15}/\textit{1.65}
    \\
    Upsampling
    & \textit{7.75}/\textit{1.58}
    & 7.51/\textit{1.84}
    & \textit{6.96}/\textit{1.80}
    & 8.95/1.79
    \\
    Both
    & \textbf{7.54}/\textbf{1.55}
    & \textbf{7.10}/\textbf{1.78}
    & \textbf{6.91}/\textbf{1.74}
    & \textbf{7.97}/\textbf{1.61}
    \\ \bottomrule
    \toprule
    (c) Model & SF-TF & \!\!SM-TM\!\! & SF-TM & SM-TF
    \\ \midrule
    V1
    & 7.91/1.60
    & 7.85/2.21
    & 7.56/2.04
    & 8.45/1.74
    \\
    V2
    & \textit{7.66}/1.56
    & \textbf{7.07}/\textit{1.81}
    & \textit{6.96}/\textbf{1.74}
    & \textit{8.07}/\textit{1.63}
    \\
    V3
    & \textbf{7.54}/\textit{1.55}
    & \textit{7.10}/\textbf{1.78}
    & \textbf{6.91}/\textbf{1.74}
    & \textbf{7.97}/\textbf{1.61}
    \\
    V2U
    & 7.69/\textbf{1.53}
    & 7.75/1.95
    & 8.11/2.33
    & 9.43/1.76
    \\ \bottomrule
  \end{tabularx}
  }
  \vspace{-4mm}
\end{table}

\smallskip\noindent\textbf{Comparison among CycleGAN-VCs.}
We analyzed the performance difference among the CycleGAN-VCs.
In addition to CycleGAN-VC (\textit{V1}), -VC2 (\textit{V2}), and -VC3 (\textit{V3}), we examined CycleGAN-VC2 with U-net~\cite{ORonnebergerMICCAI2015} (\textit{V2U}), which might be useful as an alternative to TFAN to propagate the source information to converted features.
Table~\ref{tab:objective_evaluation}(c) summarizes the results.
\textit{V2} and \textit{V3} generally have the best or second-best scores.
Among the two, \textit{V3} showed a better performance in most cases in terms of both metrics.
Furthermore, we show the qualitative and subjective comparisons in Figure~\ref{fig:mel-spectrogram} and Section~\ref{subsec:subjective_evaluation}, respectively.

\begin{figure}[t]
  \centering
  \includegraphics[width=0.99\columnwidth]{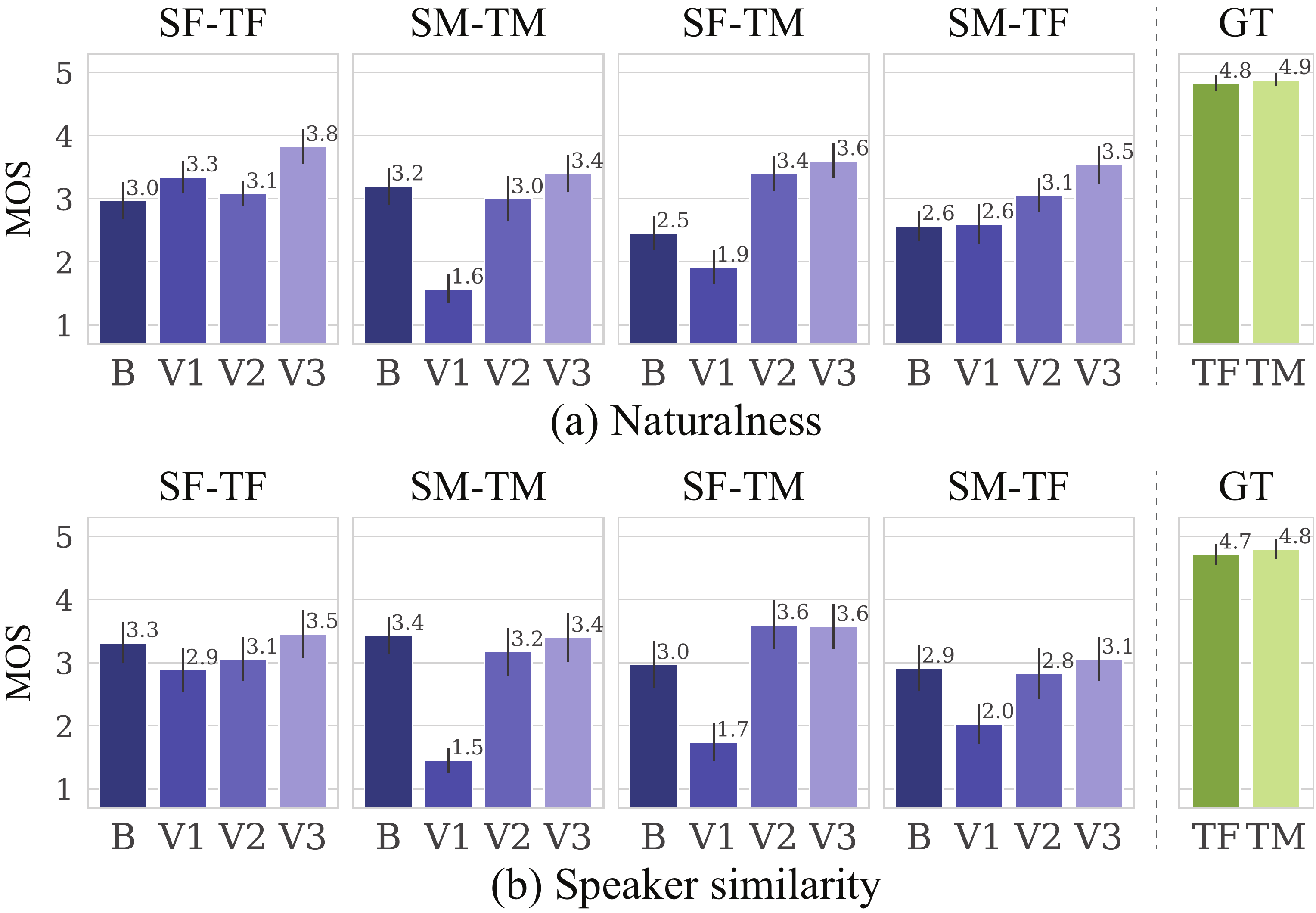}
  \vspace{-2mm}
  \caption{
    MOS for (a) naturalness and (b) speaker similarity with 95\% confidence intervals.
  }
  \vspace{-5mm}
  \label{fig:mos}
\end{figure}

\subsection{Subjective evaluation}
\label{subsec:subjective_evaluation}

We conducted listening tests to assess the applicability of CycleGAN-VCs in mel-spectrogram conversion.
We compared four models: CycleGAN-VC2 with mel-cepstrum conversion (the current best and benchmark model; denoted by \textit{B}) and CycleGAN-VC, -VC2, and -VC3 with mel-spectrogram conversion (denoted by \textit{V1}, \textit{V2}, and \textit{V3}, respectively).
To measure the naturalness and speaker similarity, we conducted mean opinion score (MOS) tests,\footnote{In the naturalness test, 1 = bad, 2 = poor, 3 = fair, 4 = good, and 5 = excellent. In the speaker similarity test, 1 = very different, 2 = moderately different, 3 = fair, 4 = moderately similar, and 5 = very similar.} where we included the target ground-truth speech (\textit{GT}) as anchor samples.
In both tests, all evaluation data (35 utterances) were used for evaluation, and each utterance was evaluated once.
In the speaker similarity test, we paired the converted speech and target speech (of which utterance contents were different) in random order.
Nine and 11 listeners participated in the naturalness and speaker similarity tests, respectively.
Audio samples are available online.\footnoteref{foot:samples}

Figure~\ref{fig:mos} shows the results.
Our main findings are as follows.
\textit{(1) Comparison between CycleGAN-VC2 for mel-cepstrum (B) and that for mel-spectrogram (V2).}
\textit{V2} is particularly effective for inter-gender VC (SF-TM and SM-TF) in terms of naturalness, and for the speaker similarity, the results are case-dependent.
This indicates that the direct application to the mel-spectrogram (i.e., \textit{V2}) is not necessarily reasonable.
\textit{(2) Comparison among CycleGAN-VCs (V1, V2, and V3).}
In most cases, the performance improves in terms of both metrics as the version increases.
This confirms the utility of \textit{V2} and \textit{V3} compared with the previous versions.
\textit{(3) Effectiveness of CycleGAN-VC3 (V3).}
\textit{V3} has a better or competitive performance compared with the other models.
This indicates the potential use of \textit{V3} as a new benchmark method in lieu of \textit{B}.

\section{Conclusions}
\label{sec:conclusions}

Although CycleGAN-VCs are widely used as benchmark methods, their feasibility for mel-spectrogram conversion is not sufficiently examined.
Therefore, following our examination, we proposed CycleGAN-VC3, an improvement of CycleGAN-VC2 that incorporates TFAN.
Using this module, we can convert the mel-spectrogram while adaptively reflecting the source mel-spectrogram.
The experimental results indicate the potential of using CycleGAN-VC3 as a new benchmark method to replace CycleGAN-VC2.
Application of TFAN to advanced tasks, such as multi-domain VC~\cite{HKameokaSLT2018,TKanekoIS2019,HKameokaArXiv2020} and application-side VC~\cite{AKainSC2007,KNakamuraSC2012,ZInanogluSC2009,TTodaTASLP2012,TKanekoIS2017b}, remains an interesting future direction.

\section{Acknowledgements}
\label{sec:acknowledgements}

This work was supported by JSPS KAKENHI 17H01763 and JST CREST Grant Number JPMJCR19A3, Japan.

\clearpage
\bibliographystyle{IEEEtran}
\bibliography{refs}

\end{document}